\documentclass[pre,twocolumn]{revtex4}
\usepackage{epsfig}
\usepackage{bm}
\usepackage{amsmath}
\usepackage{hyperref}
\usepackage{breakurl}

\newcommand{\pd}{\partial}

\begin{document}

\title{Stratification versus turbulence in complex convection}

\author{A. Bershadskii}

\affiliation{
ICAR, P.O. Box 31155, Jerusalem 91000, Israel
}

\begin{abstract}

  Natural convection is usually complicated by additional factors such as rotation, shear, radiative transfer,  compressibility, and electromagnetic fields (in the case of electro-conductive fluids). It is shown, using results of numerical simulations and measurements in the atmospheric boundary layer and solar photosphere that strong stratification can transform turbulence into deterministic chaos with exponential spectral decay of kinetic energy. When the stratification becomes weaker the deterministic chaos is replaced by the distributed chaos with stretched exponential spectral decay controlled by the second or third moments of the helicity distribution.  

\end{abstract}

\maketitle

\section{Introduction}

  The interaction of strong stratification with turbulence is one of the least understood aspects of fluid dynamics. This interaction is especially important  for atmospheric boundary layer \cite{ma} and for solar convection flows \cite{js},\cite{Kap}. In these cases, consideration is complicated by additional factors such as rotation, shear, radiative transfer,  compressibility, and electromagnetic fields. \\
  
   It is observed that at strong stable stratification the turbulence becomes intermittent, wave-like and is usually not in equilibrium with the complex nonturbulent motions (see for reviews Refs. \cite{ma},\cite{sb}-\cite{g}). It also seems to be important that even at the very stable conditions, local circulations can be generated by the weak initial and boundary disturbances, often appearing simultaneously on the multiple scales \cite{ma}, which indicates deterministic chaos with its sensitivity to the initial and boundary conditions. Interaction of the strong stratification with turbulence under unstable conditions (typical for dynamics in the vicinity of solar surface - a boundary between solar convection zone and photosphere)) is much less investigated \cite{Kap},\cite{lei}.\\

     The turbulence is restored (via the distributed chaos) with stratification decreasing, and the process of the restoration can be rather informative both for understanding the turbulence itself and the interaction between stratification and turbulence.  Recent progress in observations (measurements) and numerical simulations can be very helpful.\\
     
     In Section II appearance of the deterministic chaos with the exponential spectral decay in the atmospheric boundary layer at strong stable stratification is studied using the results of a numerical simulation. In Section III relevance of the adiabatic helical invariants for the inertial range of scales has been discussed. In Section IV the distributed chaos notion based on the helical invariants has been introduced and compared with the results of measurements in a rather stable atmospheric surface layer (over a lake). In Section V observations of the kinetic energy spectra in the solar photosphere are discussed. In Section VI results of a direct numerical simulation of anelastic convection have been used in order to understand the observational results discussed in the previous section. In Section VII unstable magnetohydrodynamic convection in the solar convection zone and atmosphere has been considered using the notion of distributed chaos and results of direct numerical simulations. In Section VIII the near-neutral case has been briefly discussed.

\section{Stably stratified Ekman boundary layer}

  The flow in an atmospheric (Ekman) boundary layer can be driven by a steady pressure gradient (a geostrophic forcing) with a homogeneous rotation. A constant temperature imposed at the bottom of the boundary layer (which is lower than the temperature outside the boundary layer) together with an initial temperature profile generate buoyancy effects. In the Boussinesq approximation dynamics of this flow can be described by the system of equations \cite{sb} 
 $$ 
 \frac{\partial {\bf u}}{\partial t} + ({\bf u} \cdot \nabla) {\bf u}  =  -\frac{\nabla p^{\star}}{\rho_\infty} + \nu \nabla^2 {\bf u}  +{\bf F}  \eqno{(1)}
$$
$$
{\bf F} = g \Phi {\bf e}_z +2{\boldsymbol \Omega} \times ({\bf G} -{\bf u})  \eqno{(2)}
$$
$$
\frac{\partial \Phi}{\partial t} + ({\bf u} \cdot \nabla) \Phi  =   \kappa \nabla^2 \Phi, \eqno{(3)}
$$
$$
\nabla P = -2 \rho_{\infty}{\boldsymbol \Omega} \times {\bf G} + g \rho_{\infty}{\bf e}_z  \eqno{(4)}
$$
$$
p^{\star} = p + \frac{1}{2}\rho_{\infty} {\bf u}^2      \eqno{(5)}
$$
$$
\Phi = (\theta -\theta_{\infty}) /\theta_{\infty}   \eqno{(6)}
$$
$$
\nabla \cdot {\bf u} =  0 \eqno{(7)}
$$
where $\theta$ is the local potential temperature, $ \theta_{\infty}$ is the reference potential temperature; ${\bf u}$ is the velocity, $p^{\star}$ is the modified perturbation pressure, $p$ is the deviation of the pressure from the mean pressure $P$, ${\boldsymbol \Omega}$ is the angular velocity of the Earth, ${\bf G}$ is geostrophic wind, ${\bf e}_z$ is a unit vector (along the gravitational
acceleration), $g$ is the value of the gravitational acceleration,  $\nu$ and $\kappa$ are the viscosity and thermal diffusivity, $\rho_{\infty}$ is a reference density.\\

  In the Ref. \cite{sb} the system Eqs. (1-7) was numerically solved with periodic horizontal boundary conditions. On the smooth horizontal bottom boundary the no-slip and impermeability boundary conditions were taken, whereas at top of the domain the stress-free and impermeability boundary conditions were taken together with zero heat flux. \\
  
    Figure 1 shows the power spectrum of the streamwise velocity fluctuations obtained at this numerical simulation for stable stratification case at $z/\delta_t =1.1$, where $\delta_t$ is the depth of the neutral-case turbulent Ekman layer. The spectral data were taken from Fig. 1b of the Ref. \cite{sb} ($\eta$ is the Kolmogorov scale). The dashed curve in Fig. 1 is drawn to indicate the exponential spectral decay
 $$
 E(k) \propto \exp-(k/k_c)   \eqno{(8)}   
 $$ 
 The dotted arrow indicates the position of the scale $k_c$.\\
  
   The exponentially decaying spectra, both in the frequency and wavenumber domains, are typical for the deterministic chaos rather than for fully developed turbulence  \cite{oh}-\cite{kds}. It seems that for $z/\delta_t=1.1$ the stable stratification is sufficiently strong to suppress the turbulence and transform it into deterministic chaos (see also Ref. \cite{ma}). Deeper in the boundary layer the stratification is weakening and more complex dynamics can appear. In order to take into account this phenomenon let us consider some adiabatic invariants of the system Eqs. (1-7).\\
   
 \section{Adiabatic helical invariants}  
     
In the inviscid approximation ($\nu=0$) the dynamic equation for the mean helicity is 
$$
\frac{d\langle h \rangle}{dt}  = 2\langle {\boldsymbol \omega}\cdot {\bf F}  \rangle \eqno{(9)} 
$$ 
where the vorticity - ${\boldsymbol \omega} = \nabla \times {\bf u}$, the helicity density - $h={\bf u}\cdot {\boldsymbol \omega}$ and $\langle...\rangle$ denotes  average over the spatial volume $V$. It follows from Eq. (9) that the mean helicity is not an inviscid invariant in this case. It is known that the mean helicity is a fundamental invariant for the Euler equations \cite{mt}. Generally, the main contribution to the correlation in the right-hand side of the Eq. (9) comes from the large-scale motion, but the correlation is quickly diminished with decreasing spatial scales due to the increasing influence of the random effects. 

  It may happen, however, that the higher moments of the helicity distribution can be still considered as the inviscid invariants \cite{mt},\cite{lt}.

  Let us following to the Refs. \cite{mt}\cite{lt} divide the considered space domain into cells with volumes $V_i$ moving with the fluid (in the Lagrangian description). Each of these volumes is bounded by surface $S_i$ with the boundary condition ${\boldsymbol \omega} \cdot {\bf n}=0$ on it. The moment of order $n$ of the helicity distribution can be then defined as 
$$
I_n = \lim_{V \rightarrow  \infty} \frac{1}{V} \sum_j H_{j}^n  \eqno{(10)}
$$
where helicity in the volume $V_i$ is
$$
H_j = \int_{V_j} h({\bf r},t) ~ d{\bf r}.  \eqno{(11)}
$$
   Due to the fast decrease of the correlation $\langle {\boldsymbol \omega}\cdot {\bf F} \rangle$ with the spatial scale the $H_j$ are still inviscid quasi-invariants for the cells with small enough spatial scales. For sufficiently randomized flow the main contribution to the sum in the right-hand side of the Eq. (10) for the high moments comes from these cells (cf. \cite{bt}).  Hence, the entire sum in the Eq. (10) for sufficiently large $n$ is an inviscid quasi-invariant despite the total helicity $I_1$ is not. For sufficiently randomized flows the $n=3$ and even $n=2$ can be considered sufficiently large for this effect ($I_2$ is the Levich-Tsinober invariant of the Euler equation \cite{lt}). Moreover, for the viscid flows the $I_n$ can be considered as adiabatic invariants for corresponding inertial ranges of scales according to the Kolmogorov-like phenomenology. It should be noted that this consideration can be also applied to the compressible fluids \cite{mt}.
   
\begin{figure} \vspace{-1.5cm}\centering
\epsfig{width=.45\textwidth,file=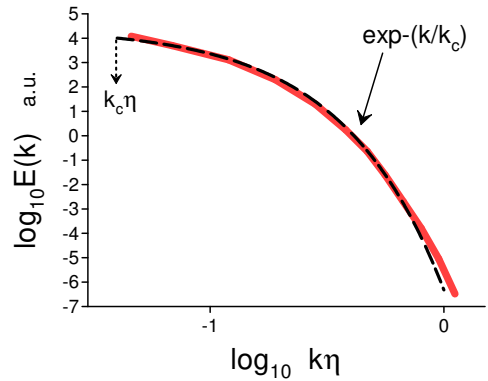} \vspace{-3.9cm}
\caption{Power spectrum of the streamwise velocity fluctuations at $z/\delta_t = 1.1$.} 
\end{figure}

\section{Distributed chaos}

When we go deeper into the boundary layer then stratification becomes weaker,  i.e. we are moving toward turbulization. In this case, the parameter $k_c$ becomes fluctuating and in order to compute the kinetic energy spectrum we should use an ensemble averaging (cf. recent Ref. \citep{b})
$$
E(k) \propto \int_0^{\infty} P(k_c) \exp -(k/k_c)dk_c \eqno{(12)}
$$    
with probability distribution $P(k_c)$. \\

  In order to find $P(k_c)$ let us use the $I_n$ adiabatic invariants of the previous Section. Actually, each of the adiabatic invariants has its attractor in the phase space. The basins of attraction of the attractors are different: the moments $I_n$ with larger $n$ have a thinner basin of attraction (intermittency).  Therefore, the domination belongs to the adiabatic invariant $I_n$ with the smallest value of $n$. Let us start from the case when $I_3$ is the adiabatic invariant with the minimal value of $n$ (i.e. the $I_2$ is not an adiabatic invariant). \\
  
    For the fluctuating $k_c$ we can estimate the corresponding characteristic velocity $u_c$ from the dimensional considerations
 $$
 u_c \propto |I_3|^{1/6} k_c^{1/2}    \eqno{(13)}
 $$
 If we assume normal (Gaussian) distribution for the characteristic velocity $u_c$ (with zero mean) \cite{my}, then it follows from the Eq. (13) that 
$$
P(k_c) \propto k_c^{-1/2} \exp-(k_c/4k_{\beta})  \eqno{(14)}
$$
where $k_{\beta}$ is a constant parameter. 
\begin{figure} \vspace{-1.5cm}\centering
\epsfig{width=.47\textwidth,file=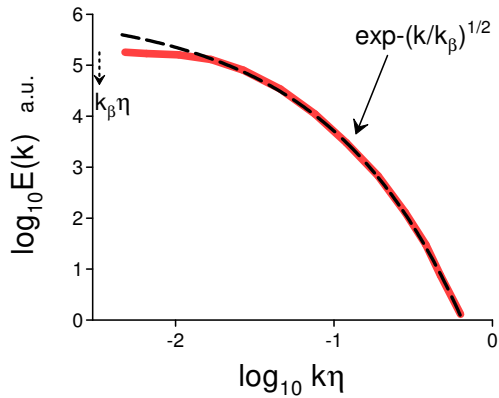} \vspace{-3.7cm}
\caption{Power spectrum of the streamwise velocity fluctuations at $z/\delta_t = 0.068$.} 
\end{figure}
\begin{figure} \vspace{-1.5cm}\centering
\epsfig{width=.45\textwidth,file=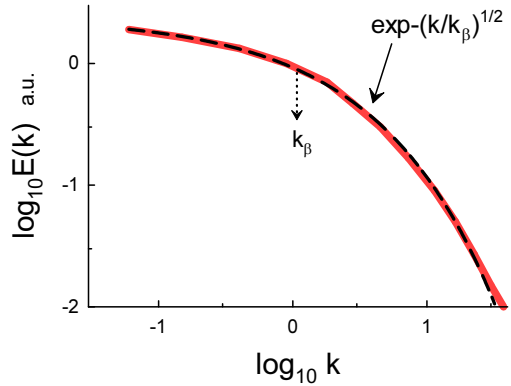} \vspace{-4.3cm}
\caption{Power spectrum of the vertical velocity fluctuations measured in a rather stable atmospheric surface layer (over a lake).} 
\end{figure}
\begin{figure} \vspace{-0.5cm}\centering
\epsfig{width=.48\textwidth,file=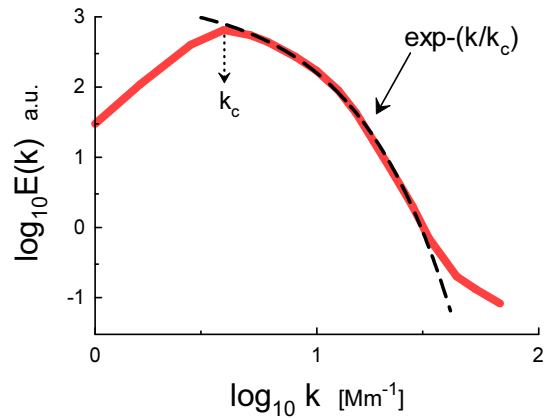} \vspace{-5.4cm}
\caption{Power spectrum of the vertical (Doppler) velocity field fluctuations in the photosphere. The spectral data were taken from the Ref. \cite{Cha}.} 
\end{figure}

    Substituting the Eq. (14) into the Eq. (12) we obtain
$$
E(k) \propto \exp-(k/k_{\beta})^{1/2}  \eqno{(15)}
$$

   It should be noted that even in the case when the odd moments of the helicity distribution are equal to zero (due to a global symmetry) this consideration can be still applied with certain modifications. The local symmetry breaking can result in the appearance of helicity cells with opposite signs of helicity so that the global helicity and the odd moments are zero (see, for instance, Ref. \cite{levich} and references therein). Let us denote the helicity of the cells with positive helicity $H_j^{+}$ and with negative helicity $H_j^{-}$, and 
$$
I_3^{\pm} = \lim_{V \rightarrow  \infty} \frac{1}{V} \sum_j [H_{j}^{\pm}]^3  
$$ 
where the summation is made for the positive (or negative) helicity cells only.  

   Due to the global symmetry $I_3 = I_3^{+} + I_3^{-} =0$ and consequently $I_3^{+} = - I_3^{-}$. Applying the same consideration as in the previous section one can conclude that $|I_3^{\pm}|$ can be still a finite (non-zero) ideal quasi-invariant and the estimate Eq. (13) can be replaced by the estimate
 $$
 u_c \propto |I_3^{\pm}|^{1/6} k_c^{1/2},    
 $$
and the kinetic energy spectrum has the same form Eq. (15) in the case of the global symmetry as well.\\

  Figure 2 shows the power spectrum of the streamwise velocity fluctuations obtained at this numerical simulation for stable stratification case at $z/\delta_t =0.068$, where $\delta_t$ is the depth of the neutral-case turbulent Ekman layer. The spectral data were taken from Fig. 1b of the Ref. \cite{sb} ($\eta$ is the Kolmogorov scale). The dashed curve in Fig. 2 is drawn to indicate the stretched exponential spectral decay Eq. (15). The dotted arrow indicates the position of the scale $k_{\beta}$.\\
  
    Figure 3 shows the power spectrum of the vertical velocity fluctuations measured in a rather stable atmospheric surface layer (over a lake). The spectral data were taken from Fig. 1e of Ref. \cite{lkb} (the wavenumber was normalized by $z$, see also Table 1 of the Ref. \cite{lkb}).  The dashed curve in Fig. 3 is drawn to indicate the stretched exponential spectral decay Eq. (15). The dotted arrow indicates the position of the scale $k_{\beta}$.

\section{Solar photosphere} 

\begin{figure} \vspace{-1.5cm}\centering
\epsfig{width=.45\textwidth,file=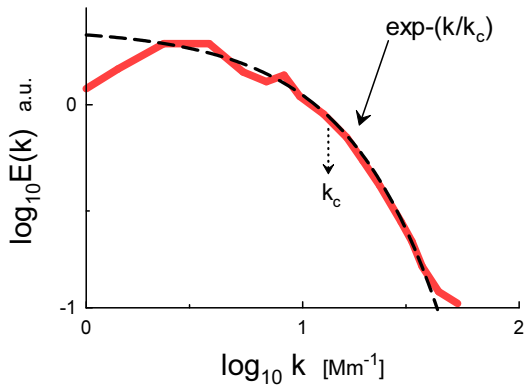} \vspace{-4.86cm}
\caption{  Power spectrum of the horizontal velocity field fluctuations in the photosphere. The spectral data were taken from the Ref. \cite{Kit}.} 
\end{figure}
\begin{figure} \vspace{-1.3cm}\centering
\epsfig{width=.45\textwidth,file=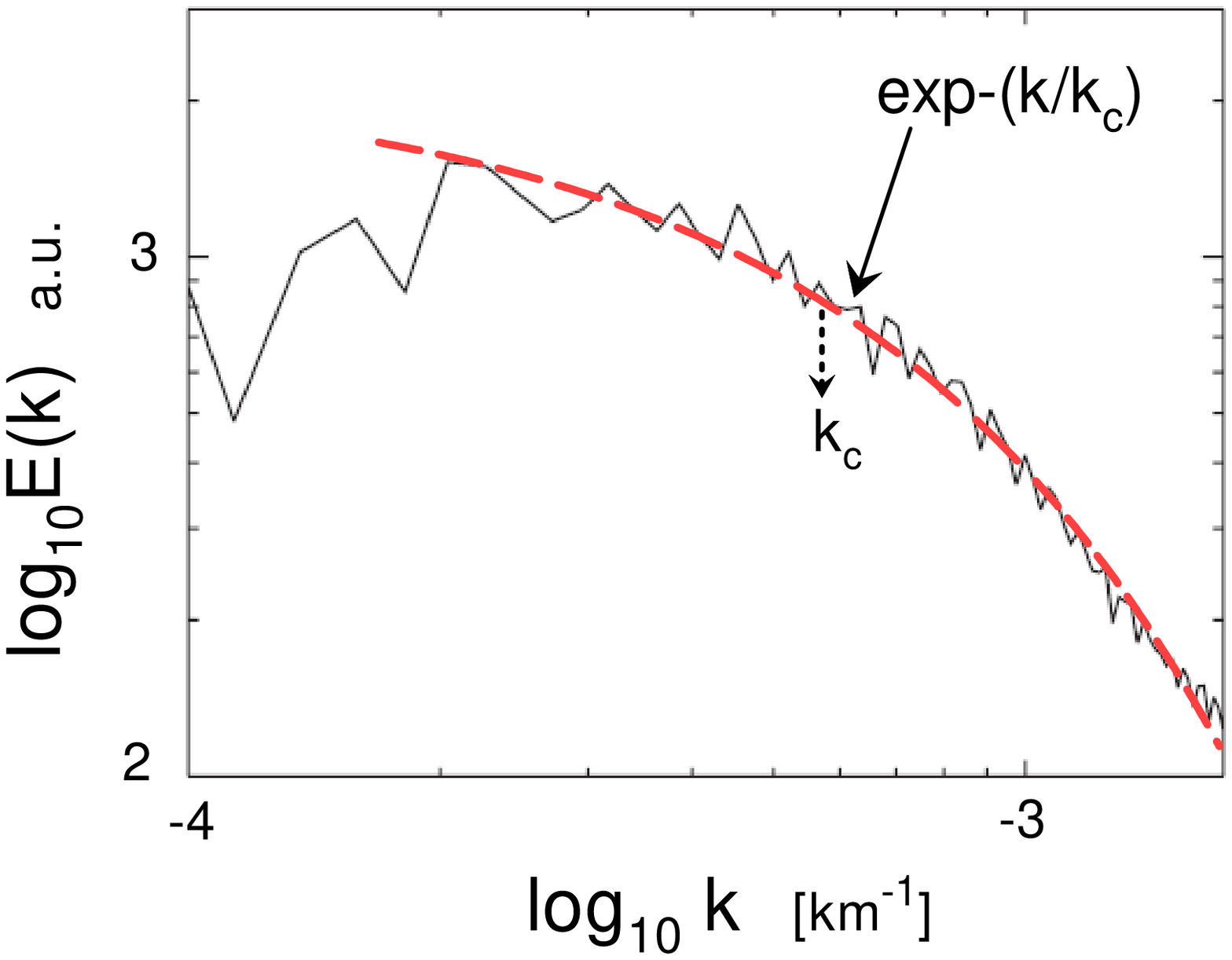} \vspace{-4.05cm}
\caption{Power spectrum of the horizontal velocity field fluctuations in the photosphere. The spectral data were taken from  the Ref. \cite{Mat}.} 
\end{figure}
\begin{figure} \vspace{-0.5cm}\centering
\epsfig{width=.45\textwidth,file=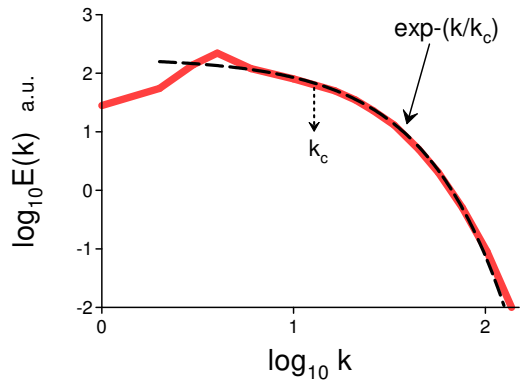} \vspace{-3.7cm}
\caption{Kinetic energy spectrum at $z/d=0.1$ (strong stratified case).} 
\end{figure}

  It is commonly believed that plasma in the solar convection zone and in the solar atmosphere should be in fully developed turbulent motion. 
  In this case, it is not sufficient to know the large scale (mean) dynamics but it is necessary to know also statistics of the velocity field fluctuations (see Refs. \cite{js},\citep{Han1} for recent reviews). \\
 
    Figure 4, for instance, shows a spatial (wavenumber) power spectrum of the vertical (Doppler) velocity field fluctuations in the photosphere obtained using the spectropolarimetric (the SUNRISE/IMaX spectropolarimeter) observations. The spectral data were taken from Fig. 3 of Ref. \cite{Cha}. 
 The dashed curve is drawn to indicate the exponential spectral decay Eq. (8). The dotted arrow indicates the position of the scale $k_c$. It should be noted that in this case the rate of the spectral decay (the scale $k_c$) is controlled by the large-scale coherent structures - the peak in the spectrum.\\

   Figure 5 shows a spatial (wavenumber) power spectrum of the reconstructed by the LCT ( local correlation tracking) horizontal velocity field fluctuations in the photosphere obtained using infrared TiO observations  (Big Bear Solar Observatory). The spectral data were taken from Fig. 3b of the Ref. \cite{Kit}. 
 The dashed curve is drawn to indicate the exponential spectral decay Eq. (8). The dotted arrow indicates the position of the scale $k_c$. \\ 
 
   This observational result is supported by the observations with the Solar Optical Telescope onboard of the Hinode spacecraft \cite{Mat}.   Figure 6 shows a spatial (wavenumber) power spectrum of the reconstructed by LCT horizontal velocity field fluctuations in the photosphere (the spectral data were taken from Fig. 5 of the Ref. \cite{Mat}).\\
   
    It should be noted that the spectra were obtained for quiet regions of the solar photosphere.\\
 
    These observational results are surprising, because as it was already mentioned above, the exponentially decaying spectra, both in the frequency and wavenumber domains, are typical for the deterministic chaos  \cite{oh}-\cite{mm} rather than for fully developed turbulence  (the Refs. \cite{Han2},\cite{Bek} can be also interesting in this respect).\\

 \section{Anelastic convection in the photosphere}   
    
     In order to understand this situation let us remind that strong stable stratification can transform turbulence into deterministic chaos with the characteristic exponential spectrum. For the photospheric heights, where the observations reported in Figs. 4-6 were made, the stratification is strong and stable (see, for instance, the Ref. \cite{csm} and references therein). \\
     
\begin{figure} \vspace{-1cm}\centering
\epsfig{width=.45\textwidth,file=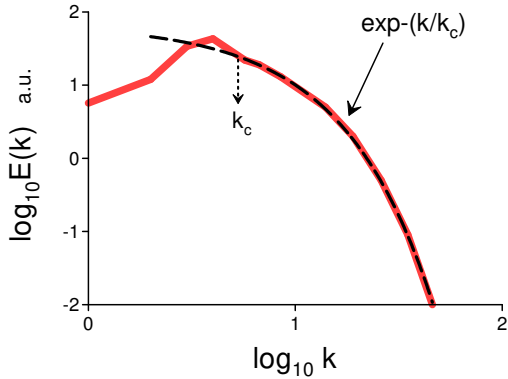} \vspace{-3.67cm}
\caption{Kinetic energy spectrum at $z/d=0.9$ (strong stratified case).} 
\end{figure}
    
\begin{figure} \vspace{-0.7cm}\centering
\epsfig{width=.45\textwidth,file=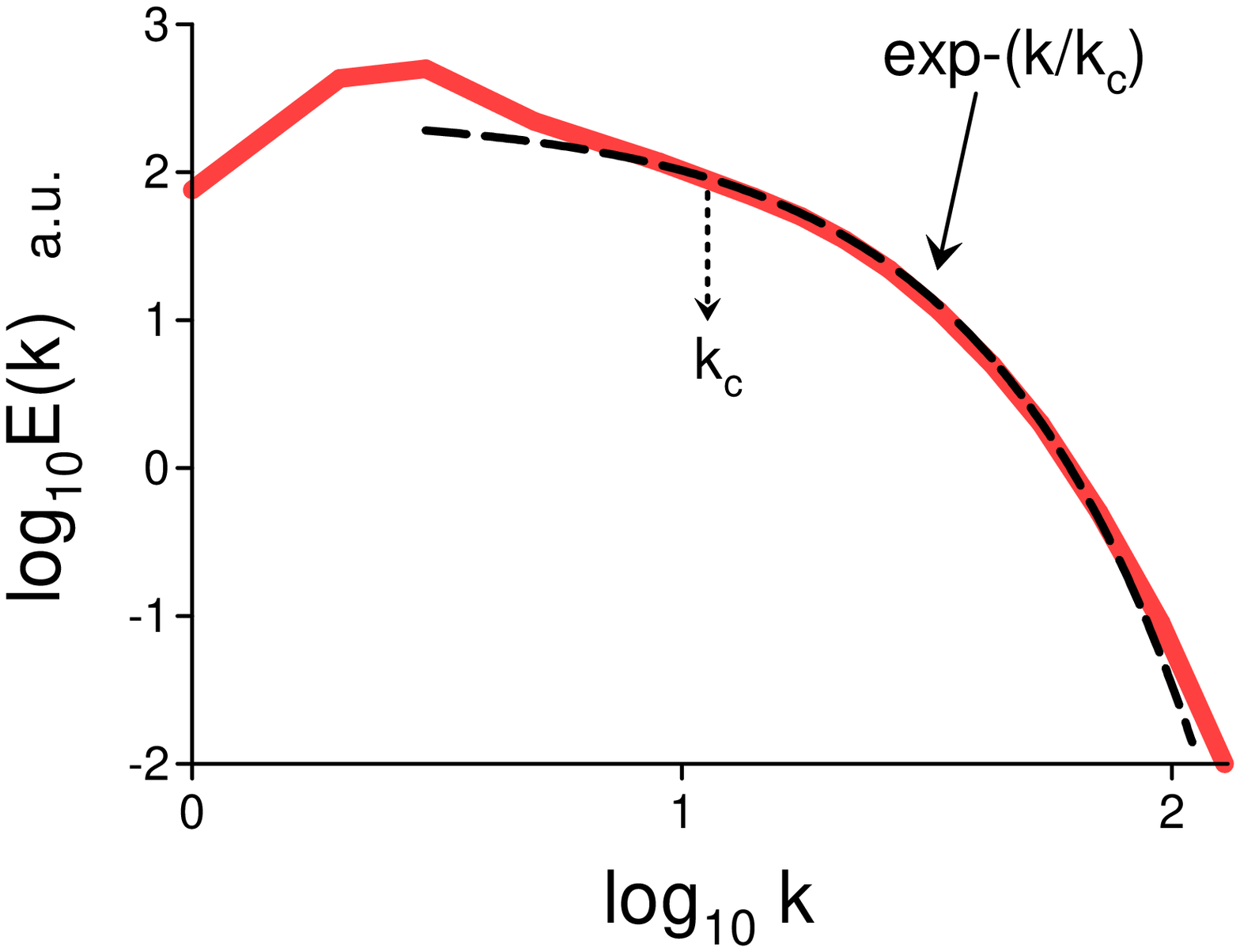} \vspace{-4cm}
\caption{Kinetic energy spectrum at $z/d=0.1$ (mildly stratified case).} 
\end{figure}
\begin{figure} \vspace{-0.55cm}\centering
\epsfig{width=.45\textwidth,file=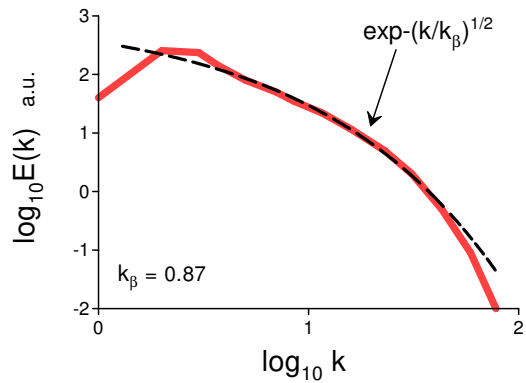} \vspace{-3.7cm}
\caption{Kinetic energy spectrum at $z/d=0.9$ (mildly stratified case).} 
\end{figure}

     In the recent Ref. \cite{kes} an attempt to advance beyond the Boussinesq approximation was made using direct numerical simulations in the anelastic approximation with filtering out sound waves and taking into account the strong stratification typical for the photosphere. The system of equations used in the Ref. \cite{kes} is
     
$$
\frac{\partial {\bf u}}{\partial t} + {\bf u} \cdot \nabla {\bf u} = -\nabla \left( \frac{p}{\bar{\rho}}\right) +{\bf F} + Pr ~{\bf D}   \eqno{(16)}
$$
$$
{\bf F} =  Ra Pr ~ s  {\bf e_z},   \eqno{(17)}
$$
$$
{\bf D} = \frac{1}{\bar{\rho}} \left( \nabla^2 {\bf u} + \frac{1}{3} \nabla \left( \nabla \cdot {\bf u} \right)  \right) ,    \eqno{(18)}
$$
$$
\nabla.\left(\bar{\rho} {\bf u}\right)  = 0 ,  \eqno{(19)}
$$
$$
\frac{\partial s}{\partial t} + {\bf u} \cdot \nabla s = \frac{ u_z}{1+\theta z} 
+ \frac{1}{\bar{\rho}} 
\left( \nabla^2 s +\frac{ \theta}{ \bar{T} } \frac{\partial s}{\partial z} \right)  
- \frac{\theta \,Q}{Ra \bar{T}\bar{\rho}},   \eqno{(20)}
$$
$$
 Q =   2 \sum\limits_{i=1}^3 \left(\frac{\partial u_i}{\partial x_i}\right)^2
  + \frac{2}{3}\left(\nabla \cdot {\bf u} \right)^2 \\
  + \sum\limits_{i<j}^3 \left(\frac{\partial u_i}{\partial x_j} +
    \frac{\partial u_j}{\partial x_i} \right)^2.    \eqno{(21)}
$$

 The square spatial domain was taken periodic in the horizontal directions. At the bottom ($z = 0$) and top ($z=d$) of the liquid layer's boundaries the impermeable and stress-free boundary conditions were taken for the velocity field and constant entropy ($s=0$) was assumed at these boundaries.\\ 

   The Prandtl number - $Pr$, and Rayleigh number - $Ra$, were defined by:
$$
  Pr=\frac{\nu}{\kappa}, 
  \qquad
  Ra = \frac{g  d^3 \varepsilon}{\nu_r \kappa_r}.  \eqno{(22)}
$$
where $\nu$ and $\kappa$ are the turbulent kinematic viscosity and thermal diffusivity respectively ($\nu_r$ and  $\kappa_r$ are their values at the bottom of the liquid layer $z =0$), the parameter $\varepsilon$ characterizes departure from adiabaticity (see Ref. \cite{kes} for more detail setup).\\

\begin{figure} \vspace{-1.7cm}\centering
\epsfig{width=.45\textwidth,file=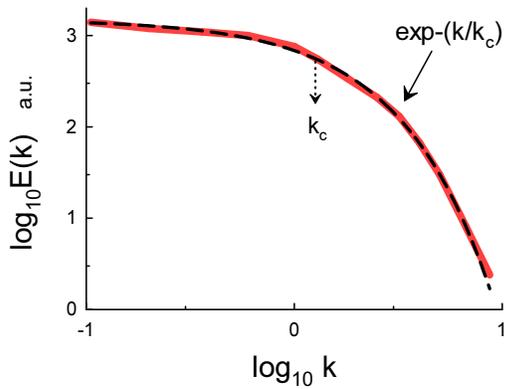} \vspace{-4.4cm}
\caption{Power spectrum of velocity fluctuations near solar surface ($z = -0.6$Mm). The spectral data were taken from  the Ref. \cite{Kap}.} 
\end{figure}

\begin{figure} \vspace{-0.5cm}\centering
\epsfig{width=.48\textwidth,file=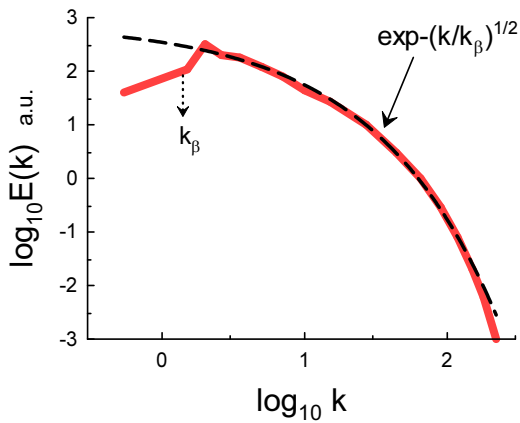} \vspace{-5.55cm}
\caption{ Power spectrum of velocity fluctuations deeper in the convection zone ($z = -18$Mm).} 
\end{figure}
\begin{figure} \vspace{-1.5cm}\centering
\epsfig{width=.45\textwidth,file=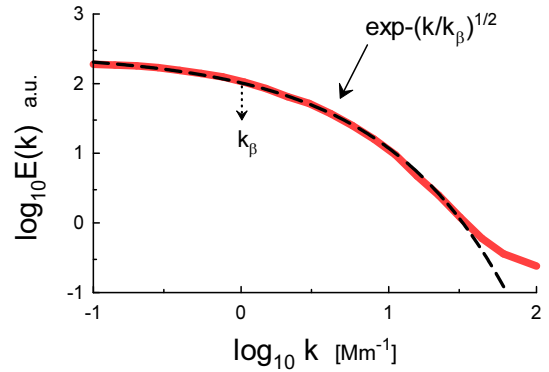} \vspace{-4.6cm}
\caption{Kinetic energy spectrum at $z=0$Mm.  The spectral data were taken from \cite{Mart}.} 
\end{figure}

  The Boussinesq approximation  is a particular case of the Eqs. (16-21) for $\theta =0$.\\

   Figures 7 and 8 show kinetic energy spectra against the horizontal wavenumber for a strong stratification case at $z/d=0.1$ and $z/d = 0.9$, respectively. The spectral data were taken from Fig. 8b of the Ref. \cite{kes}. The dashed curves are drawn to indicate the exponential spectral decay Eq. (8). \\
   
   Figure 9 shows the kinetic energy spectrum for a mildly stratified case at $z/d=0.1$. The spectral data were taken from Fig. 8a of the Ref. \cite{kes}. The dashed curve is drawn to indicate the exponential spectral decay Eq. (8). 
   
    Figure 10 shows the kinetic energy spectrum for the mildly stratified case at $z/d=0.9$. The spectral data were taken from Fig. 8a of the Ref. \cite{kes}. The dashed curve is drawn to indicate the stretched exponential spectral decay Eq. (15) (cf. Fig. 8 obtained for strong stratification).

\section{Unstable MHD convection in solar dynamics}

       Figure 11 shows the power spectrum of velocity fluctuations observed near the solar surface ($z = -0.6$Mm) in a numerical simulation reported in \cite{Kap}. The solar surface at $z=0$ separates the solar convection zone ($z < 0$) and photosphere ($z > 0$). The spectral data were taken from Fig. 13 (top panel) of the Ref. \cite{Kap}. In this simulation, a nearly isothermal thin cooling layer was added at the top of the simulated convection zone in order to maximize the stratification (the density of plasma is changing exponentially in the region). The dashed curve is drawn to indicate the exponential spectral decay Eq. (8).\\
   
     In the simulation reported in the \cite{Kap} the compressible equations
$$
\frac{\pd \bf A}{\pd t} = {\bf u}\times{\bf B} - \eta \mu_0 {\bf J},  \eqno{(23)}
$$
$$
\frac{\pd \rho}{\pd t} = -\frac{1}{\xi^2} \bf\nabla\cdot(\rho \bf{u}),  \eqno{(24)}
$$
$$
\rho\frac{D\bf{u}}{Dt} = \left[-\bf\nabla p + \bf\nabla \cdot(2\nu\rho{\bf S}) \right] +{\bf F},  \eqno{(25)}
$$
$$
{\bf F} =\rho \bf{g} +{\bf J} \times {\bf B},   \eqno{(26)}
$$
$$
T\frac{D s}{Dt} = \frac{1}{\rho}\left[{\bf\nabla} \cdot (K {\bf\nabla} T + \chi \rho T {\bf\nabla} s) + \mu_0\eta {\bf J}^2\right] + 2\nu {\bf S}^2 + \Gamma ,   \eqno{(27)}
$$
were solved, with $D/Dt = \pd /\pd t + ({\bf u} \cdot {\bf\nabla}$), $\bf{u}$ as the velocity filed, $p$ as the
pressure, ${\bf A}$ as the magnetic vector potential, ${\bf B} ={\bf B}_0 + {\bf \nabla}\times{\bf A}$ as the magnetic field, ${\bf B}_0$ as the imposed magnetic field, ${\bf J} =\mu_0^{-1}\bf\nabla\times{\bf B}$ as the current density, $\mu_0$ as the vacuum
permeability, $\eta$ as the magnetic diffusivity, $\rho$ as the plasma density, $\xi$ as the reduction factor of the sound speed,  ${\bf g}=-g\hat{\bf e}_z=$ const as the gravitational acceleration,  $K$ as the radiative heat conductivity, $\chi$ as the subgrid scale heat conductivity, $s$ as the specific entropy, $T$ as the temperature and $\Gamma$ as the cooling at the surface. It is assumed that the plasma obeys an ideal gas law $p=(\gamma-1)\rho e$, with $\gamma=c_{\rm P}/c_{\rm V}=5/3$ and $e=c_{\rm V} T$ as the internal energy. The rate-of-strain tensor (traceless)
$\mbox{\boldmath ${\sf S}$}$ is
$$
{\sf S}_{ij} = \frac{1}{2} (U_{i,j}+U_{j,i}) - \frac{1}{3} \delta_{ij} {\bf \nabla} \cdot \bf{U}.   \eqno{(28)}
$$
The Smagorinsky viscosity
$\nu=(C_k\Delta)^2\sqrt{\mbox{\boldmath ${\sf S}$}^2}$, with $\Delta$
as the filtering scale (the grid spacing), and $C_k=0.35$. \\

 The bottom and top boundaries of the simulated layer domain were taken stress-free and impenetrable. The temperature was fixed at the top boundary. For more details on the numerical simulation setup see the Ref. \citep{Kap}. \\
 
    When we go deeper into the convection zone then, according to the Ref. \cite{Kap}, stratification becomes weaker and the effective Reynolds number becomes larger, i.e. we are moving toward turbulization, more precisely toward the distributed chaos (see Section IV). \\
    
    In this case, the Eq. (9) should be replaced by the equation
 $$
\frac{d\langle h \rangle}{dt}  = \frac{2}{\rho}\langle {\boldsymbol \omega}\cdot {\bf F}  \rangle \eqno{(29)} 
$$    
where ${\bf F}$ is given by the Eq. (26), but the consideration of Sections III and IV remains valid.  \\
  
  Figure 12 shows the power spectrum of velocity fluctuations observed in the same numerical simulation \cite{Kap} at $z =-18$Mm, i.e. deeper in the convection zone. The spectral data were taken from Fig. 13 (top panel) of the Ref. \cite{Kap} ($k$ is the horizontal wavenumber normalized by the density scale height). The dashed curve in Fig. 12 is drawn in order to indicate correspondence to the Eq. (15).\\

\begin{figure} \vspace{-1.8cm}\centering
\epsfig{width=.45\textwidth,file=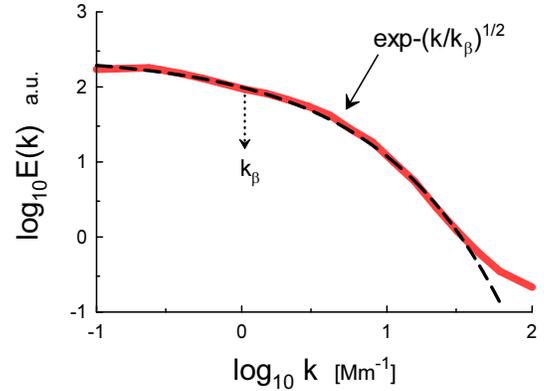} \vspace{-4.1cm}
\caption{As in Fig. 13 but for  $z=-1$Mm.} 
\end{figure}
\begin{figure} \vspace{-0.5cm}\centering
\epsfig{width=.45\textwidth,file=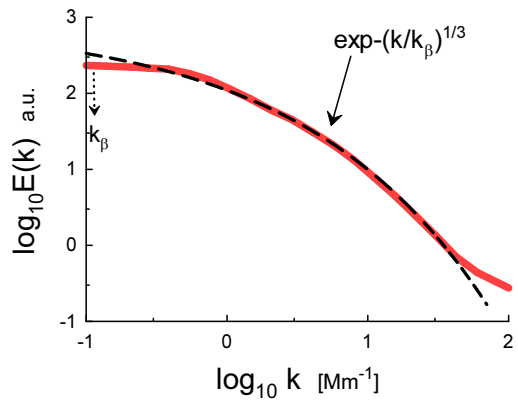} \vspace{-4.33cm}
\caption{As in Fig. 13 but for  $z=0.81$Mm.} 
\end{figure}

 In the numerical simulation reported in the Ref. \cite{Kap} special efforts were made in order to investigate the influence of strong stratification on the plasma dynamics near the solar surface. In the most of other numerical simulations of the solar convection there were no such special efforts and it is interesting to look at their results for both the convection zone and for solar atmosphere. \\
 
 Figure 13 shows the power spectrum of velocity fluctuations observed at the solar surface ($z = 0$Mm) in a numerical simulation reported in recent Ref. \cite{Mart} and obtained using a radiative magnetohydrodynamic simulation of the solar atmosphere with non-LTE and non-grey radiative transfer with thermal conduction along the magnetic fields. The spectral data were taken from Fig. 6 of the Ref. \cite{Mart}. The dashed curve in Fig. 13 is drawn in order to indicate correspondence to the Eq. (15). In this simulation no special efforts were made to take into account the strong stratification near the solar surface and one can see that the obtained at $z=0$Mm kinetic energy spectrum is similar to that shown in Fig. 12 (the helical distributed chaos with the stretched exponential spectrum Eq. (15)) and not to the Fig. 11 (deterministic chaos with exponential spectrum Eq. (8)).\\
       
         Figure 14 shows the kinetic energy at $z=-1$Mm (in the terms of the Ref. \cite{Mart} it means the convective zone). The spectral data were taken from Fig. 6 of the Ref. \cite{Mart}. The dashed curve in Fig. 14 is drawn in order to indicate correspondence to the Eq. (15).

\section{Near-neutral case}       

 As it was already mentioned at certain conditions the second-order moment of the helicity distribution $I_2$ also can be considered as an adiabatic invariant in an inertial range of scales. In this case, the estimate Eq. (13) should be replaced by the estimate
$$
 u_c \propto |I_2|^{1/4} k_c^{1/4}    \eqno{(30)}
 $$    
also obtained from the dimensional considerations. \\

The above used straightforward calculation of the ensemble-averaged spectrum is impossible in this case and we will use an asymptotic method. Let us generalize the stretched exponential spectrum Eq. (15) as
 $$
E(k) \propto \int_0^{\infty} P(k_c) \exp -(k/k_c)dk_c \propto \exp-(k/k_{\beta})^{\beta} \eqno{(31)}
$$  
where $k_{\beta}$ and $\beta$ are constants. The probability distribution $P(k_c)$ can be estimated from the Eq. (31) in the asymptotic of large $k_c$ as 
\cite{jon} 
$$
P(k_c) \propto k_c^{-1 + \beta/[2(1-\beta)]}~\exp(-\gamma k_c^{\beta/(1-\beta)}) \eqno{(32)}
$$     
where $\gamma$ is a constant.\\

\begin{figure} \vspace{-1.5cm}\centering
\epsfig{width=.45\textwidth,file=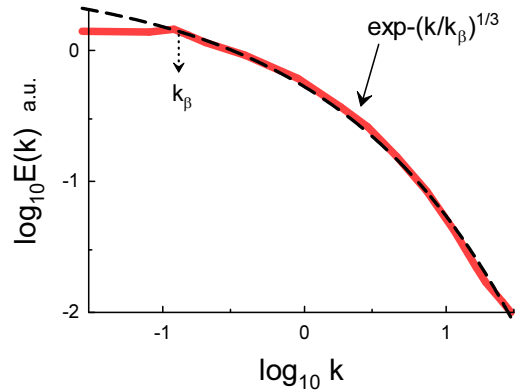} \vspace{-4.5cm}
\caption{Power spectrum of the vertical velocity fluctuations measured in a near-neutral atmospheric surface layer (over a lake).} 
\end{figure}

   Let us generalize the scaling relationships  Eqs. (13) and (30)  as   
$$
u_c \propto  k_c^{\alpha}   \eqno{(33)}
$$

 If $u_c$ has Gaussian distribution a relationship between $\alpha$ and $\beta$ follows immediately from the Eqs. (32) and (33)
$$
\beta = \frac{2\alpha}{1+2\alpha}  \eqno{(34)}
$$
 
   Substituting $\alpha = 1/4$ from the Eq. (30) into the Eq. (34) we obtain
 $$
E(k) \propto \exp-(k/k_{\beta})^{1/3}  \eqno{(35)}
$$       
 instead of the Eq. (15). \\
 
    Figure 15 shows the kinetic energy spectrum obtained at $z=0.81$Mm in the numerical simulation Ref. \cite{Mart}. This value of $z$ corresponds to the low-chromosphere in the terms of the numerical simulation (the layer of the chromosphere just atop the photosphere is nearly homogeneous, see for a review Ref. \cite{rut}). The spectral data were taken from Fig. 6 of the same Ref. \cite{Mart}. The dashed curve in Fig. 15 is drawn in order to indicate correspondence to the Eq. (35).\\

  Figure 16 shows the power spectrum of the vertical velocity fluctuations measured in an atmospheric surface layer (over a lake) at near-neutral conditions. The spectral data were taken from the Fig. 1a of the Ref. \cite{lkb} (the wavenumber was normalized by $z$, see also Table 1 of the Ref. \cite{lkb}).  The dashed curve in Fig. 16 is drawn to indicate the stretched exponential spectral decay Eq. (35) (cf. Fig 3). The dotted arrow indicates the position of the scale $k_{\beta}$. One can see that in this case the rate of the spectral decay (the scale $k_{\beta}$) is controlled by the large-scale coherent structures - the peak in the spectrum (cf., for instance, Ref. \cite{sa}).\\ 
   
    Figure 16, corresponding to the near-neutral conditions, can be also helpful for understanding the situation shown in Fig. 15. The relaxation of stratification results in the decreasing of the parameter $\beta$ in the stretched exponential spectral decay Eq. (31).
  
 \section{Acknowledgement}

I thank  A. Brandenburg, P. J. K\"apyl\"a   and J. Martinez-Sykora for helpful additional information about their numerical simulations. I am also grateful to E. Levich and K. R. Sreenivasan for sending to me their papers and numerous discussions.


\begin{thebibliography}{99}
\bibitem{ma} L. Mahrt, Annu. Rev. Fluid Mech., {\bf 46} 23 (2014)
\bibitem{js} J. Schumacher and K.R. Sreenivasan, Rev. Mod. Phys., in press (2020)
\bibitem{Kap} P.J. Kapyla, A. Brandenburg, N. Kleeorin, M.J. Kapyla, I. Rogachevskii, A\&A, 588, A150 (2016)
\bibitem{sb} S.K. Shahand and E. Bou-Zeid, J. Fluid Mech., {\bf 760}, 494 (2014)
\bibitem{fw} H.J.S. Fernando and J.C. Weil, Bull. Am. Meteorol. Soc., {\bf 91}, 1475 (2010)
\bibitem{gsa} B. Galperin, S. Sukoriansky and P. Anderson, Atmos. Sci. Lett., {\bf 8} 65 (2007)
\bibitem{g} C. Gibson, J. Mar. Sys., {\bf 21} 147 (1999)
\bibitem{lei} P. Leitner, B. Lemmerer, A. Hanslmeier, T. Zaqarashvili, A. Veronig, H. Grimm-Strele, H.J. Muthsam, Astrophys. Space Sci., {\bf 362}, 181 (2017)
\bibitem{mt} H.K. Moffatt and A. Tsinober, Annu. Rev. Fluid Mech., {\bf 24}, 281 (1992)
\bibitem{lt} E. Levich and A. Tsinober, Phys. Lett. A {\bf 93}, 293 (1983)
\bibitem{bt} A. Bershadskii and A. Tsinober,  Phys. Rev. E, {\bf 48}, 282 (1993)
\bibitem{oh} N. Ohtomo, K. Tokiwano, Y. Tanaka et. al., J. Phys. Soc.
Jpn., {\bf 64}, 1104 (1995)
\bibitem{sig} D.E. Sigeti, Phys. Rev. E, {\bf 52}, 2443 (1995)
\bibitem{f} J.D. Farmer, Physica D, {\bf 4}, 366 (1982).
\bibitem{fm} U. Frisch and R. Morf, Phys. Rev., {\bf 23}, 2673 (1981)
\bibitem{mm} J. E. Maggs and G. J. Morales, Phys. Rev. Lett., {\bf 107},
185003 (2011); Phys. Rev. E {\bf 86}, 015401(R) (2012); Plasma Phys. Control.
Fusion, {\bf 54}, 124041 (2012)
\bibitem{kds} S. Khurshid, D.A. Donzis and K.R. Sreenivasan, Phys. Rev. Fluids, {\bf 3}, 082601(R) (2018)
\bibitem{b} A. Bershadskii, Res. Notes AAS, {\bf 4}, 10 (2020)
\bibitem{my} A. S. Monin, A. M. Yaglom, Statistical Fluid Mechanics, Vol. II: Mechanics of Turbulence (Dover Pub. NY, 2007)
\bibitem{levich} E. Levich, Concepts of Physics, {\bf VI}, 239 (2009)
\bibitem{lkb} D. Li, G. G. Katul, and E. Bou-Zeid, Boundary-Layer Meteorol., {\bf 157}, 1 (2015)
\bibitem{Han1} S.M. Hanasoge, H. Hotta and K.R. Sreenivasan, Sci. Adv., {\bf 6},  eaba9639 (2020)
\bibitem{Cha} L.Y.Chaouche, F. Moreno-Insertis and J.A. Bonet, A\&A, {\bf 563}, A93 (2014)
\bibitem{Kit} I.N. Kitiashvili, V.I. Abramenko P.R. Goode, A.G. Kosovichev, S.K. Lele, N.N. Mansour A.A. Wray and V.B. Yurchyshyn,  Phys. Scr., {\bf  T155},  014025 (2013)
\bibitem{Mat} B. Matsumoto and P. Kitai, ApJ. Lett., {\bf 716}, L19 (2010)
\bibitem{Han2} S.M. Hanasoge, T.L. Duvall Jr. and K.R. Sreenivasan
Proc. Natl. Acad. Sci. U.S.A., 109, 11928 (2012)
\bibitem{Bek} Y. Bekki, H. Hotta and T. Yokoyama, ApJ, {\bf 851}, 74 (2017)
\bibitem{csm} M.C.M. Cheung, M. Schüssler and F. Moreno-Insertis, A\&A, {\bf 461}, 1163 (2007)
\bibitem{kes} M. Kessar, D. Hughes, E. Kersale, K.A. Mizerski, and S.M. Tobias, ApJ, {\bf 874} 103 (2019)
\bibitem{Mart} J. Martinez-Sykora, V.H. Hansteen, B. Gudiksen, M. Carlsson, B. De Pontieu and M. Gosic, ApJ, {\bf 878}, 40 (2019)
\bibitem{jon} D.C. Johnston, Phys. Rev. B, {\bf 74}, 184430 (2006)
\bibitem{rut} R.J. Rutten, Phil. Trans. R. Soc. A, {\bf 370}, 3129 (2012)
\bibitem{sa} S. Salesky, W. Anderson, J. Fluid Mech., {\bf 856}, 135 (2018)

\end{thebibliography}
\end{document}